\documentclass[aps,prc,showpacs,reprint,superscriptaddress]{revtex4-1}
\usepackage[OT4]{fontenc}

\usepackage{graphicx}   

\begin{document}

\title{Proton spectroscopy of $^{48}$Ni, $^{46}$Fe, and $^{44}$Cr}

\date{\today}

\author{M.~Pomorski}
\affiliation{Faculty of Physics, University of Warsaw, 00-681 Warsaw, Poland}
\author{M.~Pf\"utzner}
\email{pfutzner@fuw.edu.pl}
\affiliation{Faculty of Physics, University of Warsaw, 00-681 Warsaw, Poland}
\author{W.~Dominik}
\affiliation{Faculty of Physics, University of Warsaw, 00-681 Warsaw, Poland}
\author{R.~Grzywacz}
\affiliation{Department of Physics and Astronomy, University of Tennessee, Knoxville, TN 37996, USA}
\affiliation{Physics Division, Oak Ridge National Laboratory, Oak Ridge, TN 37831, USA}
\author{A.~Stolz}
\affiliation{National Superconducting Cyclotron Laboratory, Michigan State University, East Lansing, MI 48824, USA}
\author{T.~Baumann}
\affiliation{National Superconducting Cyclotron Laboratory, Michigan State University, East Lansing, MI 48824, USA}
\author{J.S.~Berryman}
\affiliation{National Superconducting Cyclotron Laboratory, Michigan State University, East Lansing, MI 48824, USA}
\author{H.~Czyrkowski}
\affiliation{Faculty of Physics, University of Warsaw, 00-681 Warsaw, Poland}
\author{R.~D\k{a}browski}
\affiliation{Faculty of Physics, University of Warsaw, 00-681 Warsaw, Poland}
\author{A.~Fija{\l}kowska}
\affiliation{Faculty of Physics, University of Warsaw, 00-681 Warsaw, Poland}
\author{T.~Ginter}
\affiliation{National Superconducting Cyclotron Laboratory, Michigan State University, East Lansing, MI 48824, USA}
\author{J.~Johnson}
\affiliation{Physics Division, Oak Ridge National Laboratory, Oak Ridge, TN 37831, USA}
\author{G.~Kami\'nski}
\affiliation{Institute of Nuclear Physics PAN, 31-342 Cracow, Poland}
\affiliation{Joint Institute for Nuclear Research, 141980 Dubna, Moscow Region, Russia}
\author{N.~Larson}
\affiliation{National Superconducting Cyclotron Laboratory, Michigan State University, East Lansing, MI 48824, USA}
\affiliation{Department of Chemistry, Michigan State University, East Lansing, MI 48824, USA}
\author{S.N.~Liddick}
\affiliation{National Superconducting Cyclotron Laboratory, Michigan State University, East Lansing, MI 48824, USA}
\affiliation{Department of Chemistry, Michigan State University, East Lansing, MI 48824, USA}
\author{M.~Madurga}
\affiliation{Department of Physics and Astronomy, University of Tennessee, Knoxville, TN 37996, USA}
\author{C.~Mazzocchi}
\affiliation{Faculty of Physics, University of Warsaw, 00-681 Warsaw, Poland}
\author{S.~Mianowski}
\affiliation{Faculty of Physics, University of Warsaw, 00-681 Warsaw, Poland}
\author{K.~Miernik}
\affiliation{Physics Division, Oak Ridge National Laboratory, Oak Ridge, TN 37831, USA}
\affiliation{Faculty of Physics, University of Warsaw, 00-681 Warsaw, Poland}
\author{D.~Miller}
\affiliation{Department of Physics and Astronomy, University of Tennessee, Knoxville, TN 37996, USA}
\author{S.~Paulauskas}
\affiliation{Department of Physics and Astronomy, University of Tennessee, Knoxville, TN 37996, USA}
\author{J.~Pereira}
\affiliation{National Superconducting Cyclotron Laboratory, Michigan State University, East Lansing, MI 48824, USA}
\author{K.P.~Rykaczewski}
\affiliation{Physics Division, Oak Ridge National Laboratory, Oak Ridge, TN 37831, USA}
\author{S.~Suchyta}
\affiliation{National Superconducting Cyclotron Laboratory, Michigan State University, East Lansing, MI 48824, USA}
\affiliation{Department of Chemistry, Michigan State University, East Lansing, MI 48824, USA}

\begin{abstract}
Results of decay spectroscopy on nuclei in vicinity of the doubly magic $^{48}$Ni are presented. The measurements were performed with a Time Projection Chamber with optical readout which records tracks of ions and protons in the gaseous volume.
Six decays of $^{48}$Ni, including four events of two-proton ground-state radioactivity were recorded. An advanced reconstruction procedure
yielded the 2p decay energy for $^{48}$Ni of $Q_{2p}=1.29(4)$~MeV.
In addition, the energy spectra of $\beta$-delayed protons emitted in the decays of $^{44}$Cr and $^{46}$Fe, as well as half-lives and branching ratios were determined.
The results were found to be consistent with the previous
measurements made with Si detectors. A new proton line in the decay of $^{44}$Cr corresponding to the decay energy of 760 keV is reported. The first evidence for the $\beta 2p$ decay of $^{46}$Fe, based on one clear event, is shown.
\end{abstract}



\maketitle

\section{Introduction}
\label{chap:intro}

The quest to reach the limits of nuclear existence and to learn properties
of nuclides at these limits is one of most important topics in the present-day
low-energy nuclear physics. The progress in this field is largely driven by
recent advances of experimental techniques allowing efficient production, separation
and detection of very exotic nuclei, located far from the $\beta$ stability
and characterized by extreme proton-to-neutron imbalance. Although the neutron-deficient
side of the nuclidic chart is much better explored than the neutron-rich frontier,
there are still a lot of unsurveyed areas on this chart and open questions concerning
nuclei at and beyond the proton drip-line. The nuclear properties
in this region are shaped by the interplay between large $\beta$-decay $Q$ values,
low or negative proton separation energies, and the confining effects of
the Coulomb barrier. The resulting characteristic phenomena include a variety
of $\beta$-delayed particle emission channels, proton radioactivity and two-proton
radioactivity \cite{PfutznerRMP,BorgeNP152,PfutznerNP152,BlankBorge}.

The latter process, discovered 12 years ago \cite{Pfutzner2002, Giovinazzo2002} is still
not well known. Its mechanism is not fully understood and its potential to
reveal nuclear-structure information is not firmly established yet. Up to now,
the simultaneous two-proton (2p) emission from the ground state was unambiguously observed in $^{6}$Be, $^{19}$Mg, $^{45}$Fe, $^{48}$Ni, and $^{54}$Zn \cite{PfutznerRMP}. It is expected,
however, that this decay mode should be observable for almost every even-Z element
up to tellurium \cite{Olsen2013,*OlsenErrata}. In the first experiments the evidence
for 2p decay was obtained by means of arrays of Si detectors which only allowed
for a determination of the total decay energy and the decay time \cite{Pfutzner2002,Giovinazzo2002,Blank54Zn}.
To fully explore the physical information carried by the two protons, however, one has
to record their momenta separately. This requirement led to the development of new types
of detectors capable of recording tracks of charged particles in a gaseous medium,
based on the Time Projection Chamber (TPC) principle. One such device \cite{BlankTPC}
provided the first direct evidence for the 2p decay of $^{45}$Fe \cite{Giovinazzo2007}
and $^{54}$Zn \cite{Ascher2011}. In another detector, developed at the University of Warsaw,
a novel concept of optical readout was applied to a drift chamber which led to
the Optical Time Projection Chamber (OTPC) \cite{MiernikNIM}. The OTPC detector was
successfully used to measure the first full proton-proton correlation picture
for the 2p decay of $^{45}$Fe \cite{Miernik2007, MiernikEPJ}. This experiment
revealed the three-body character of the process and provided the first evidence
for the sensitivity of the 2p correlation pattern to the angular momentum
composition of the initial wave function. In addition, the OTPC detector was
instrumental in the discovery of the $\beta$-delayed three-proton ($\beta 3p$)
emission in the case of $^{45}$Fe \cite{MiernikCr2007} and $^{43}$Cr \cite{Pomorski2010}.

Recently, the OTPC detector was used to study the decay of the extremely neutron-deficient
($T_z = -4$) and presumably doubly-magic $^{48}$Ni.
The main result of this work was the first observation of the 2p decay of $^{48}$Ni.
The preliminary results on the decay of $^{48}$Ni were published in
Refs. \cite{Pomorski2011,PomorskiAPP,PomorskiPROCON}.
Here we present the results of the full and final analysis of this experiment.
An improved track
reconstruction procedure was used to accurately determine the energies of the
detected particles.
In addition to the 2p decay of $^{48}$Ni, the $\beta$-delayed protons emitted
in the decays of $^{46}$Fe and $^{44}$Cr were recorded. Decays of both these nuclei
were studied before by means of implantation in a stack of Si-detectors by Dossat et al. \cite{Dossat}.
The comparison of our data with those from Ref. \cite{Dossat} provides
a consistency check for our algorithms of the data analysis. It is used also to
point out advantages of the TPC technique over Si detectors in the charged particle
spectroscopy of exotic nuclei, especially in detecting low-energy protons.
Such comparison demonstrates the complementarity of both techniques.

Section II of this paper presents the experimental details concerning the production, separation, and in-flight identification of the ions of interest.
The OTPC system is described with the focus on modifications and improvements with
respect to the detector used in the 2p spectroscopy
of $^{45}$Fe \cite{Miernik2007,MiernikEPJ}. In Sec. III the main steps of the data
analysis are presented, in particular the procedure for the proton track reconstruction.
The main results are listed and discussed in Sec. IV which is followed by conclusions in Sec. V.

\section{Experimental Technique}
\label{chap:Exp_tech}
\subsection{Production and identification of ions}
\label{chap:Production&ID}
The experiment was carried out at the National Superconducting Cyclotron Laboratory (NSCL). The ions were produced in a fragmentation reaction by bombarding a $580 \textrm{ mg/cm}^2$ natural nickel target with a $^{58}\textrm{Ni}$ beam with an energy of $160$ MeV/nucleon. A rotating target assembly was developed for this experiment by Oak Ridge National Laboratory and the University of Tennessee. In the course of the experiment the target was run at speeds up to 900 RPM and withstood beam currents up to 40 pnA. The ions of interest were separated from contaminants using the A1900 fragment separator \cite{A1900} in the achromatic setting with two aluminum degraders mounted in the I1 and I2 focal planes. The degraders had thickness of $193$ mg/cm$^2$ and $302$ mg/cm$^2$, respectively. Selected ions were transferred to the S2 vault where the OTPC detector system was placed. The average time of flight of ions from the target to the detector, calculated with the LISE code \cite{LISE}, was about $500$ ns.

\begin{figure}
\includegraphics[width = 1\columnwidth]{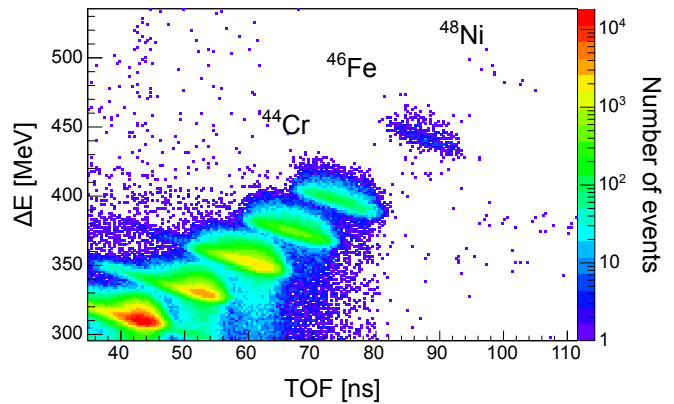}
\caption{\label{fig:NSCLID} (Color on-line) The identification spectrum of all ions arriving to the S2 vault, as collected by the A1900 standard acquisition system.}
\end{figure}

Each fragment arriving at the detector was identified using the time-of-flight and energy-loss technique. The time-of-flight (TOF) was measured between a plastic scintillator, positioned at the middle focal plane of the A1900 separator, and a Si detector placed at the end of the beam line just before the OTPC detector. This Si detector also provided the energy loss ($\Delta E$) data. The average rate of ions at detector setup in the S2 vault was about $10$ ions/s. The complete data were recorded by the standard acquisition system of the A1900 separator. The resulting identification plot is presented in Fig. \ref{fig:NSCLID}.

\subsection{The OTPC detection system}
\label{chap:OTPC}
The Optical Time Projection Chamber (OTPC) was developed at the University of Warsaw specifically to study
very rare decay modes with emission of charged particles, like 2p radioactivity.
The main concept and some details of the unit used in the study of $^{45}$Fe were given in Ref. \cite{MiernikNIM,MiernikEPJ}.
For the present experiment a new chamber was designed and produced.
Here we briefly summarize its main features.

\begin{figure}
\includegraphics[width = \columnwidth]{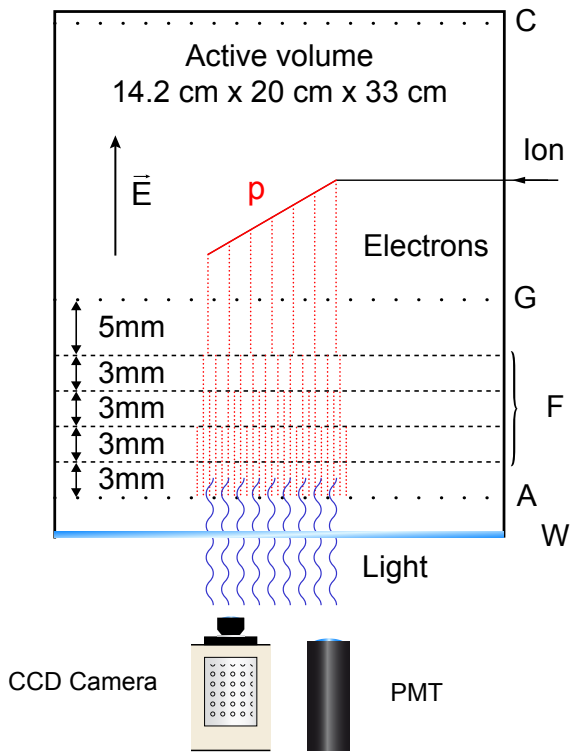}
\caption{\label{fig:OTPC} (Color on-line) Schematic view of the OTPC chamber with an
example event of $\beta$-delayed proton emission from a stopped ion. Only ionization electrons
resulting from the proton are marked. The letters on the right indicate: C - cathode,
G - gating electrode, F- set of 4 GEM foils, A - wire-mesh anode, W - window transparent to visible light.}
\end{figure}

The detector is schematically shown in Fig. \ref{fig:OTPC}. The active volume, having dimensions of $33\times20\times14.2$ cm$^3$ (depth, width and height, respectively), is filled with a gaseous mixture at atmospheric pressure. In this study a gas mixture of $49.5$\% Ar, $49.5$\% He, and $1\%\textrm{ N}_2$ was used.

Within the active volume, between the cathode and the amplification stage, a constant and uniform
electric field is maintained with the help of copper electrodes spaced evenly on the side walls.
The direction of this field is vertical and its strength was about $E=210$~V/cm.

The incoming ions enter the active volume horizontally through a kapton entrance window.
If the ion is stopped inside the active volume, its decay with emission of heavy charged
particles, like protons or $\alpha$ particles, can be registered. Primary electrons, resulting
from the gas ionization by the stopping ion and by emitted charged particles, drift with the
constant velocity $v_d$ toward the amplification stage, passing through the gating electrode.
We note that electrons emitted during $\beta$ decays generate ionization too weak to be registered by the detector.
In order to allow for the registration of both the implanted ion and the particles emitted in the decay, a gating electrode connected to a fast-switching high-voltage power supply was used. By changing the potential of this electrode, we could either block most of the primary ionization electrons or let them pass to the amplification section. These settings are referred to as the "low sensitivity" and "high sensitivity" regimes, respectively. Switching between these two settings takes about $100 \mu$s.

The signal amplification is performed using four Gas Electron Multiplier foils (GEM) \cite{GEM}. The voltage between the two sides of each foil and the voltage between the neighboring foils can be controlled individually, in order to optimize the performance of the system. The former were set in the range between $240$ and $280$ volts and were tuned during the experiment in order to maintain maximum possible gain. The voltages between GEM foils were set to $800$ V. Between the last GEM foil and the final anode electrode a high voltage of $1000$ V was supplied, causing electrons to stimulate light emission from particles of the gas mixture. At this point the electric signal is converted to light.

This light is registered with a digital camera (CCD) and a photomultiplier (PMT) connected to an oscilloscope. In this experiment a $512$x$512$ 16 bit pixel back-thinned CCD camera (Hamamatsu c9100-13) and a $100$ MS/s 14 bit per sample oscilloscope (NI PXI-5142) were used. The CCD image represents a projection of an event on the plane of GEM foils, integrated over exposure time (typically around 30~ms). The PMT trace provides the total light intensity as a function of time, which allows for a determination of the time between the implantation and the decay. In addition, the PMT signal contains the information of the event along the direction of the electric field, i.e. perpendicular to the anode plane. Moreover, if the entire track of an emitted proton is contained within the active volume, it can be reconstructed in 3D by combining data from the PMT and the CCD.

The chamber used in this experiment differs in a few key aspects from the detector used
in the study of $^{45}$Fe and described in Ref. \cite{MiernikNIM}. First, the ions enter the detection volume perpendicularly to the electric field and not diagonally, as before. Thus, contrary to
the previous case, the distance of the stopped ion to the amplification stage does not depend on the implantation depth. Even more important is that the ions do not penetrate the amplification section, which
could cause malfunctions due to large ionization. Second, the wire-mesh electrodes were replaced by the GEM foils which reached the same amplification with smaller voltages applied and resulted in much more stable working conditions. The effects of
electric discharges, which blocked the
previous detector are now practically absent. Finally, we use a new CCD camera with a better quantum efficiency.

In order to optimize the implantation depth of the ions of interest, an adjustable degrader was placed in front of the OTPC entrance window. In this experiment an 832 $\mu$m thick Al degrader was used. However, due to the large energy spread of ions coming from the A1900 separator, only about $65$\% of $^{48}$Ni ions could be stopped in the active volume of the OPTC, the rest either punched through the chamber or stopped in the
entrance window or before.

\begin{figure}
\includegraphics[width = \columnwidth]{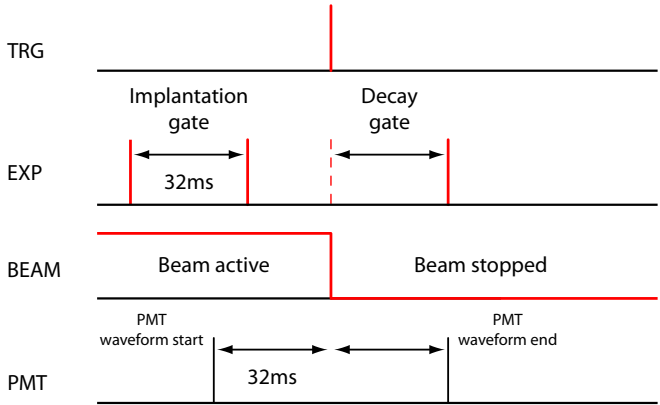}
\caption{\label{fig:Data_sequence} The sequence of events in the extended exposition mode of the OTPC operation. TRG shows the arrival time of an ion of interest, EXP marks exposures of the CCD camera, BEAM indicates the stopping of the primary beam, while PMT shows the range of the registered waveform from the photomultiplier. }
\end{figure}

The OTPC acquisition system was triggered selectively, based on the $\Delta$E-TOF information for the incoming ion. The trigger signal was activated only by ions for which both the TOF and the $\Delta E$
values exceeded certain limits. Those limits were adjusted to accept all ions of $^{48}$Ni and of $^{46}$Fe, and a small part of $^{44}$Cr ions. During the entire experiment a special "extended exposure" mode of operation was used. In this mode, while awaiting the trigger, the OTPC is kept in the "low sensitivity" regime and the CCD camera is continuously taking images with a constant exposure time (referred to as the "implantation gate"). These images are discarded unless a trigger signal arrives during the exposure. Upon the arrival of the trigger the OTPC is switched to the "high sensitivity" regime and the CCD exposure is extended by a fixed time period (referred to as the "decay gate"). At the same moment the primary beam is stopped to prevent other ions from entering the detector in the high sensitivity mode.
The signal waveform from the PMT is stored in a circular buffer of the digital oscilloscope.
The trigger determines the time span of the recorded waveform. It starts one length of the implantation gate before the trigger and spans over the entire exposure time until the end of the decay gate. This sequence of events is shown in Fig. \ref{fig:Data_sequence}.

During the experiment the implantation gate was set to $32$ ms at all times. The decay gate was set to $32$~ms for most of the time, however, some data were taken with a longer decay gate of $120$ ms.
These settings are referred to as the short and the long exposure, respectively. The PMT signal was sampled with $50$ MHz and $25$ MHz for events taken in the short and long exposure mode, respectively. The CCD camera used in this experiment could not
accept a trigger for $780~\mu$s after each implantation exposure.
This introduced a dead time of $2.4\%$ and $0.6 \%$ in the short and long
exposure settings, respectively.

\begin{figure}
\includegraphics[width = \columnwidth]{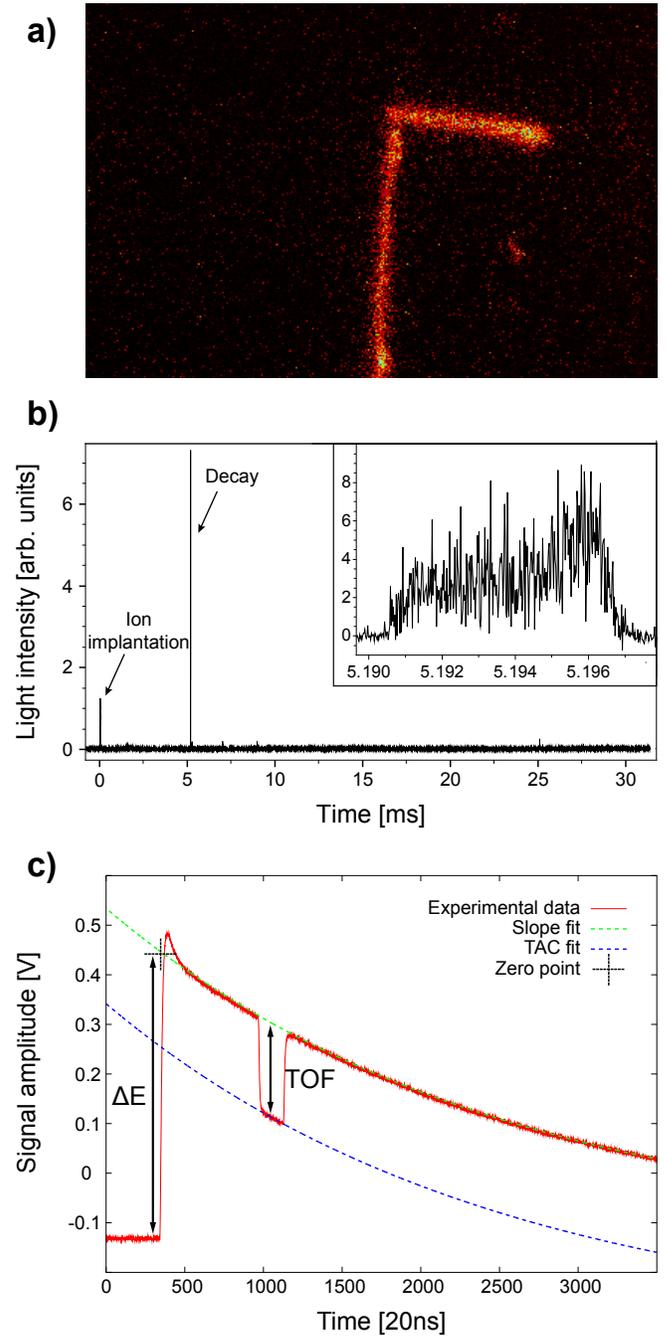}
\caption{\label{fig:EventData} (Color on-line) An example set of data recorded by the OTPC
detector for one event of $\beta$-delayed proton emission from $^{46}$Fe. a) On the CCD image
a track of an ion coming from below and a track of a proton emitted about 5 ms later are visible; b) The waveform of the PMT signal shows the sequence of events and the zoomed decay
part in the insert; c) the ID information for the ion consists of the signal from the Si detector and the superimposed signal from the TAC. The fitted curves used to extract the
corresponding values of the $\Delta E$ and the TOF are also shown.}
\end{figure}

For each event the ID information for the triggering ion was recorded.
The $\Delta E$ signal from the Si detector, pre-amplified and processed by a fast
amplifier, and the delayed signal from the time-to-amplitude converter (TAC) representing
the TOF of the ion, were combined by means of a linear summing module. Then this
signal was fed to a second channel of the oscilloscope which recorded its full waveform.
After the decay gate was closed, all collected data, comprising the CCD image,
the waveform of the PMT signal, and the waveform of the ID signals were read and stored
on a disk. Since the primary beam was switched-off for a period of about 1 s after the trigger, there was ample time for data read-out and storage before the beam was switched on again and
the OTPC was ready for another trigger.
Example data recorded for one event are displayed in Fig.\ref{fig:EventData}.

\section{Data analysis}
\label{chap:Data_analysis}

\subsection{Ion identification}
\label{chap:ID_fit}

\begin{figure}
\includegraphics[width = \columnwidth]{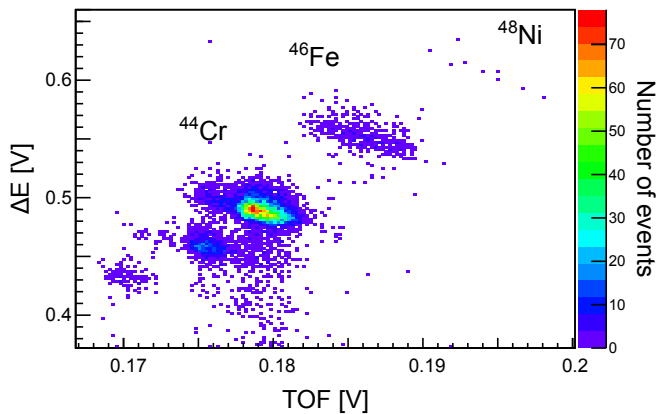}
\caption{\label{fig:IDspec} (Color on-line) The identification spectrum extracted from the
ID data recorded by the OTPC showing all ions which triggered the OTPC acquisition system.}
\end{figure}

From the ID information of each recorded event (Fig. \ref{fig:EventData}c) the identification spectrum of all ions which triggered the OTPC system was constructed. By fitting the shape of
the signals with help of the Levenberg-Marquardt algorithm, as implemented in the levmar 2.5 library \cite{levmar}, the relevant physical parameters were extracted. First, from the exponential slope of the Si signal its amplitude and thus the value of the $\Delta$E was determined.
Then, taking this slope into account, the amplitude of the TAC signal was found yielding
the value of the TOF.
The resulting ID spectrum for all recorded events is presented in Fig. \ref{fig:IDspec}.

In total 8580 events were collected, $6563$ were taken with the short exposure and $2017$ with the long exposure time. We identify $9$ events of $^{48}$Ni, $471$ events $^{46}$Fe and $5542$ events of $^{44}$Cr.

 An inspection of Fig.\ref{fig:NSCLID} reveals that the A1900 acquisition system recorded 10 events of $^{48}$Ni, thus one more than the OTPC system. The A1900 system was taking
data independently of the status of the OTPC. This one event could
have been missed by the OTPC acquisition either due to the CCD camera induced
dead-time or if it arrived when the OTPC was not collecting data, for example during
adjustments of the OTPC.

\subsection{Track reconstruction}
\label{chap:Reconstruction}

When a charged particle emitted in a decay is stopped inside the active
volume of the OTPC, the recorded data allows for a determination of the energy
and the direction of the particle's track in the 3D space. The procedure which we
developed for such a reconstruction is based essentially on comparison of
the observed track with simulations.

First, the regions from the CCD image and the PMT waveform which contain
the signal from the particle are cut out yielding two experimental
distributions for further processing.
Since the ratio between the energy deposited by a particle in the gas
and the registered amount of light was not perfectly stable,
we analyze only the shape of the distributions, ignoring the absolute amplitude.
Therefore, both distributions are normalized to yield the integral of 1.
In addition they are smoothed with a gaussian filter to reduce high frequency
noise. The parameters of this filter were kept constant and the same for all events.

We assume that the measured signal, both in the CCD image and in the
PMT waveform, is proportional to the primary ionization density which in turn is
proportional to the stopping power, $dE/dx$, of the charged particle. Using the
SRIM2013 code \cite{SRIM} we calculate the stopping power profile
along the trajectory of the particle in the OTPC gas mixture for a given initial energy
of the particle. Further, for an assumed particle emission angle, we project the calculated energy-loss profile on the anode plane (horizontal) and on the direction perpendicular to it (vertical). The length scale of the
vertical profile is expressed in the units
of time assuming the constant drift velocity $v_d$ of electrons in the OTPC chamber.
To account for the diffusion of the drifting charge,
we introduce a spread to the projected profiles by a gaussian function.
The widths characterizing the diffusion in the horizontal and the vertical
directions can be different. Finally, both simulated profiles are smoothed
in the same way as the experimental distributions and normalized to the unit integral.
The two profiles thus produced can be compared with the experimental distributions.

\begin{figure}
\includegraphics[width = \columnwidth]{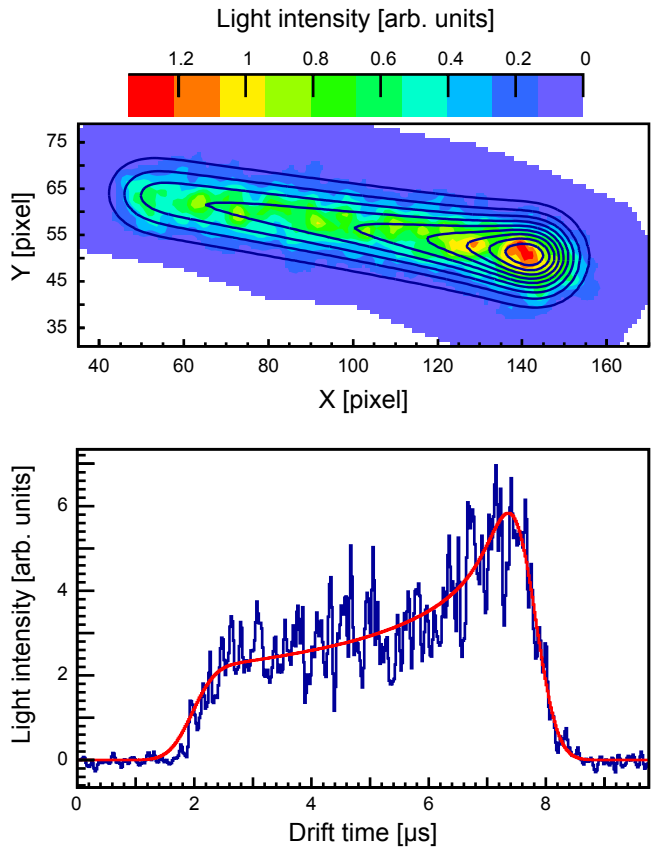}
\caption{\label{fig:1p} (Color on-line) Reconstruction of a $\beta$-delayed proton track from the
event shown in Fig. \ref{fig:EventData}. The best fitting simulation is indicated by the
contour lines on the CCD image (top) and by the red line on the PMT waveform (bottom). }
\end{figure}

\begin{figure}
\includegraphics[width = \columnwidth]{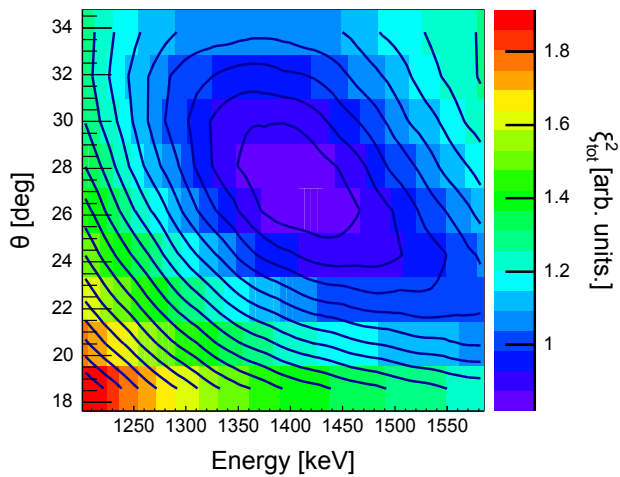}
\caption{\label{fig:xi_map} (Color on-line) The map of $\xi_{tot}^2$ values around the
minimum value for the event shown in Fig. \ref{fig:EventData} plotted as a function of the proton energy and the angle $\theta$ between the proton's track and the horizontal plane. Each point
represents the value minimized over all parameters except the energy and $\theta$. }
\end{figure}

To quantify the comparison of the simulated CCD response
with the corresponding experimental distribution we introduce the function:
\begin{equation}
	\xi^2_{CCD} = \sum_{i,j} \left[CCD_{exp}(i,j) - CCD_{sim}(i,j)\right]^2,
\end{equation}
where $CCD_{exp}(i,j)$ and $CCD_{sim}(i,j)$ are the smoothed, normalized experimental signal, and the simulated OTPC's response for the pixel coordinates $i,j$, respectively. The summing runs over all pixels of the experimental distribution. Similarly, for the PMT signal we define:
\begin{equation}
	\xi^2_{PMT} = \sum_{i} \left[PMT_{exp}(i) - PMT_{sim}(i)\right]^2,
\end{equation}
where $PMT_{exp}(i)$ and $PMT_{sim}(i)$ are the $i^{th}$ element of the experimental and the simulated PMT signal, respectively. Again the summing range covers the whole signal.
Finally, we combine both functions:
\begin{equation}
	\xi^2_{tot} = \frac{\xi^2_{CCD}}{w_{CCD}} + \frac{\xi^2_{PMT}}{w_{PMT}},
\end{equation}
where $w_{CCD}$ and $w_{PMT}$ are the weighting factors reflecting the corresponding
number of degrees of freedom. For the $w_{CCD}$ we take the length of the track on the
CCD image in pixels, and for the $w_{PMT}$ we take the number of samples in the PMT waveform.
The reconstruction of the particle track is done by running the simulations for
various values of the initial energy, emission angles and the two diffusion widths,
to find the set of parameters which minimizes the function $\xi^2_{tot}$.
To illustrate this procedure, the reconstruction results for the event
shown in Fig. \ref{fig:EventData} are shown in Figs. \ref{fig:1p} and \ref{fig:xi_map}.


The events with the simultaneous emission of two protons were reconstructed using the
same procedure. To simulate the detector response for such an event, first each proton
was simulated individually. Then, both simulations were merged assuming that the emission originated from same place and occurred the same time.
In cases where it was not clear which part of the PMT signal corresponds to which track
on the CCD image, all possible configurations were simulated and the one providing
the least $\xi^2_{tot}$ was chosen.

\subsection{Energy calibration using $^{44}$Cr }
\label{chap:calibration}

To verify the reconstruction procedure and to fine-tune the value of the electron
drift velocity $v_d$, the $\beta$-delayed protons emitted in the decay of $^{44}$Cr
were used. The spectrum of delayed protons for this case was measured by
Dossat et al. \cite{Dossat}. We were able to reconstruct 103 tracks of protons
originating from $^{44}$Cr which were fully confined in the OTPC active volume.
The resulting energy spectrum, see Fig. \ref{fig:Cr}, clearly shows pronounced
peaks. Two lines, marked in Fig. \ref{fig:Cr} as 2 and 3, correspond to the lines
found in Ref. \cite{Dossat} at 908~keV and 1384~keV, respectively.
Taking into account that we are extracting the kinectic energy of the proton
while results of Ref. \cite{Dossat} refer to the decay energy which includes
the recoil of the daughter nucleus, we do reproduce these energy values with the drift
velocity of $v_d = 6.00 (25)$ mm/$\mu$s which fixes the energy calibration.
The results for the decay of $^{44}$Cr will be discussed in more detail in Sec. IV A.

\subsection{Uncertainties}

The final uncertainties of the reconstruction procedure were estimated
by combining the inaccuracy of the $\xi^2$ minimization, and the systematic error of the
drift velocity. The total uncertainty of proton energy
was found to range from 4\% to 8\%. The angle $\theta$ of a proton track
with respect to the horizontal plane is determined with an accuracy of
about 4$^ \circ$. For the event shown in Fig. \ref{fig:EventData} the reconstruction
procedure yielded the proton kinetic energy $E_p = 1393 (50) (6)$~keV
and the track angle with respect to the horizontal plane
$\theta = 28^\circ (4^\circ) (1^\circ)$, where the first error
corresponds to the statistical uncertainty of the $\xi ^2$ minimization
and the second error reflects
the systematic uncertainty of the drift velocity.

\section{Results}

\subsection{$^{44}$Cr}

\subsubsection{Half-life and the total branching ratio}

Out of $5542$ ions identified as $^{44}$Cr by the OTPC ID procedure, 4098 were stopped
well inside the active volume of the chamber at a sufficient distance from the walls to
ensure that the emission of a delayed proton is clearly visible. In 183 events such an emission
indeed was observed. Although in many cases the emitted proton escaped the OTPC volume,
this number together with the number of well implanted ions allows for the determination
of the half-life and the total branching ratio for the $\beta$-delayed proton emission.

To extract the half-life of $^{44}$Cr the maximum likelihood method was used
combining events registered with both the short and the long exposure.
Following the procedure described in Ref. \cite{Pomorski2010} the half-life was found to be $T_{1/2} = 25^{+6}_{-4}$~ms. The reason for the large error bars is that most
of the events were collected in the short exposure mode, with the decay gate of 32 ms
being of the same order as the measured half-life.
This result agrees within 3$\sigma$ with the value reported by Dossat et al. of $T_{1/2} = (43 \pm 2)$~ms \cite{Dossat}.

In the analysis of the branching ratio one has to take into account
the fact that the $\beta$ particles are not observed in the OTPC.
Thus, an event picturing only the implanted ion
indicates that either no decay occurred within the observation time (decay gate)
or the $\beta$ decay did occur but without emission of delayed protons. Since the
half-life and the length of the decay gate are known, the maximum likelihood
method can be used to determine the branching ratio in such case \cite{Pomorski2010}.
Using the more precise half-life value measured by Dossat et al. we found
that the total branching ratio for the $\beta$-delayed proton emission by $^{44}$Cr
is $b_{\beta p} = 10(1)\%$. This is to be compared with the value reported
by Dossat et al. of 14.0(9)\% \cite{Dossat}. We note that our method is essentially
based on counting the incoming ions and the decay events, and its ultimate accuracy
is limited only by statistics. In particular, it is free of systematical errors
present in the method used by Dossat et al. \cite{Dossat}, who had to impose
an arbitrary cut on the proton energy spectrum to avoid the significant
background due to $\beta$ particles. In addition, our method does not
suffer from the uncertainty of normalization. The relative uncertainty of our result is
of the same order as that of Ref. \cite{Dossat} but the decay gate in our
measurement was not optimized for the decay of $^{44}$Cr and the number of
collected ions of $^{44}$Cr in our experiment was smaller by an order of magnitude.

\subsubsection{Energy spectrum}

\begin{figure}
\includegraphics[width = \columnwidth]{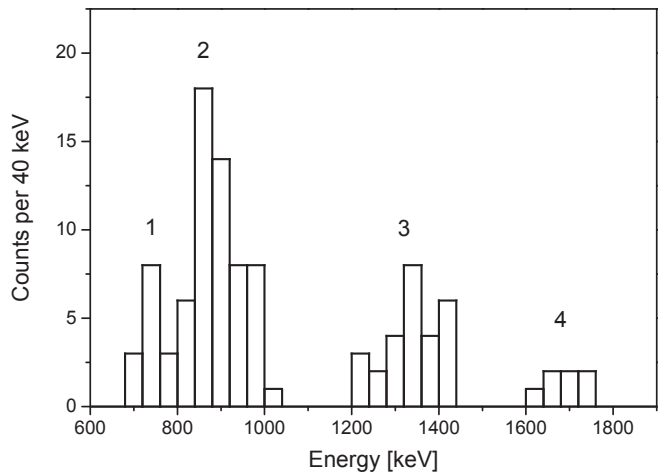}
\caption{\label{fig:Cr} Energy spectrum of $\beta$-delayed protons emitted from $^{44}\textrm{Cr}$ with four lines marked.}
\end{figure}

Using the procedure described in Sec. \ref{chap:Reconstruction}, we have reconstructed all events of $\beta p$ emission in which the full proton track was recorded. In total, $103$ decay events of $^{44}$Cr could be successfully reconstructed and the energy spectrum of the emitted protons is shown in Fig. \ref{fig:Cr}.

\begin{table}
\caption{\label{tab:ResCr}Energies of proton groups observed in the
decay of $^{44}$Cr and the corresponding branching ratios. The $E_p$
denotes the proton kinectic energy. The values for Ref. \cite{Dossat}
were recalculated from published decay energy values by correcting for
the daughter recoil. }
	\begin{ruledtabular}
		\begin{tabular}{ccccc}
            & \multicolumn{2}{c}{This work} & \multicolumn{2}{c}{Ref. \cite{Dossat}} \\
		    & $E_p$ [keV] & $I_p$ [\%] & $E_p$ [keV] & $I_p$ [\%] \\
		\hline
		$1$ & $742(24)(10)\footnotemark[1] $ & $ 0.6(2)$  &    & \\
		$2$ & $896(53)\footnotemark[2]  $    & $ 2.7(5)$  &  887(11) & 1.7(3) \\
		$3$ & $1340(62)\footnotemark[2] $     & $ 1.4(3)$ & 1353(12) & 1.1(3) \\
		$4$ & $1680(44)\footnotemark[2] $     & $ 0.5(2)$ & 1700(15) & 0.6(3) \\
		\end{tabular}
	\end{ruledtabular}
\footnotetext[1]{The first error comes from the minimization procedure, while the second reflects the uncertainty of the drift velocity.}
\footnotetext[2]{The energy value calculated as the average of events in the peak area and the error corresponds to the standard deviation of this average.}
\end{table}

The three broad structures seen in this Figure (marked as 2, 3, and 4) correspond to the
peaks reported by Dossat et al. \cite{Dossat} at the decay energy of 908 keV, 1384 keV, and 1741 keV, respectively. The widths of these peaks are larger than the energy resolution,
both in Ref. \cite{Dossat} and in this work, indicating that they are composed
of overlapping lines. In addition, our spectrum shows a narrow structure at
742 keV (marked as 1 in Fig. \ref{fig:Cr}), consistent with a single proton line.
This line has not been identified by Dossat et al. probably because of the large
$\beta$ background, see Fig. 15b of Ref. \cite{Dossat}. This illustrates the
advantage of the OTPC detector which is not sensitive to $\beta$ particles.

On the other hand, due to the limited active volume, protons of
high energy escape the OTPC detector which presents a disadvantage
if compared to an array of Si detectors. While most of the protons at 1000 keV
are fully stopped inside the OTPC, the probability to escape strongly
increases with the proton energy. At the energy of 1800 keV, the length of the proton
track in the OTPC gas mixture is about 11 cm, thus many such protons escape
the active volume and cannot be reconstructed. That is why at
about 1700 keV (peak 4 in Fig. \ref{fig:Cr}) we see only a part
of the real intensity. To correct for this effect we determined the
efficiency of stopping the entire proton track within the fiducial volume of
the detector as a function of proton energy by a Monte Carlo method.
The ranges of protons in the OTPC gas mixture were calculated by the SRIM2013 code \cite{SRIM}, the measured implantation profile of $^{44}$Cr ions in the
gas volume was taken into account, and the isotropic emission of $\beta$-delayed protons
was assumed.
Taking the total proton branching ratio and the number of counts
from Fig. \ref{fig:Cr}, and correcting for the
stopping efficiency, we have obtained the branching
ratios for individual peaks. They are presented in Table \ref{tab:ResCr}
in comparison with results of Ref. \cite{Dossat}.
As long as emitted protons are stopped within the OTPC, our method provides more
accurate values of the branching ratios as the spectrum is not affected by the
background of $\beta$ particles.

\subsection{$^{46}$Fe}

\subsubsection{Production cross-section}

Although the ion-optical setting of the A1900 separator was not optimal for $^{46}$Fe, the transmission of this nucleus was large enough for the determination of the production cross section. According to the procedure described in Ref. \cite{cross}, the cross-section is given by:
\begin{equation}
\sigma = \frac{N_{Fe}}{N_{proj}} \frac{A_t}{N_A d_t} \frac{1}{T_1 T_2},
    \label{eq:cs}
\end{equation}
where $N_{proj}$ and $N_{Fe}$ is the number of beam particles which hit the target and the number of $^{46}$Fe ions identified, respectively, $A_t$ is the molar mass of the target, $N_A$ is the Avogadro number, $d_t$ is the target thickness in g/cm$^2$, $T_{1}$ is the transmission of $^{46}$Fe from the target to the final focus of the A1900 separator, which takes into account losses in the material of the target and in the degraders, and $T_2$ represents the transmission from the A1900 final focus through the beam line to the Si detector.

We use the number of identified ions, $N_{Fe}$, from the A1900 standard identification system (see Fig. \ref{fig:NSCLID}) which does not suffer any dead-time limitations. We found $N_{Fe}=503$, while the number of projectiles, $N_{proj}$, was determined by a Faraday cup to be $N_{proj} = 8 \times 10^{16}$. The target of $d_t = 580$ mg/cm$^2$ thickness was made of natural nickel with $A_t = 58.7$~g. The transmission $T_1$ was calculated by LISE++ code \cite{LISE} using the momentum distribution according to the model of Morrisey \cite{Morrissey}, which yielded $T_1 = 0.13(6)$.
The large uncertainty of this value is dominated by the uncertainty of the shape of the momentum
distribution. It was estimated by comparing predictions of different
models of this distribution \cite{LISE}.
The transmission $T_2$ was determined experimentally to be $T_2 = 0.40(5)$. Finally, the production cross section for $^{46}$Fe in the fragmentation reaction of $^{58}$Ni beam at 160 MeV/nucleon on a natural nickel target
is $\sigma = (25 \pm 12)$~pb. This number is fairly well reproduced by the EPAX 3 parametrization which predicts $\sigma^{EPAX} = 14$~pb \cite{EPAX3}.

\subsubsection{Half-life and the total branching ratio}

In the OTPC identification spectrum 471 events of $^{46}$Fe were found. Out of this number 269 ions were stopped well inside the chamber, far enough from the walls to see the eventual emission of $\beta$-delayed particles. In 139 events such emission was observed, which allows for the determination of the half-life and the total branching ratio for $\beta$-delayed proton emission.

Using the maximum likelihood method we found that the half-life of $^{46}$Fe is
$T_{1/2} = 16.4^{+4.2}_{-2.8}$ ms, which is consistent with the value
of $T_{1/2} = 13.0(17)$ ms reported by Dossat et al. \cite{Dossat}. By combining these
two values according to the procedure described in Ref. \cite{BarlowError} we obtain
the more accurate result of $T_{1/2} = 14.0^{+1.4}_{-1.3}$ ms.

Using this combined value of the half-life and following the same procedure
as for $^{44}$Cr (sec. IV.A.1), the total branching ratio for the emission
of $\beta$-delayed protons in the decay of $^{46}$Fe is found to be
$b_{\beta p} = 66(4)\%$. This result agrees within $3\sigma$ with the value
of $b_{\beta p} = 79(4)\%$ obtained by Dossat et al. \cite{Dossat}.
Our result does not suffer from any systematical uncertainties due to the $\beta$ background.

\subsubsection{Energy spectrum}

\begin{figure}
\includegraphics[width = \columnwidth]{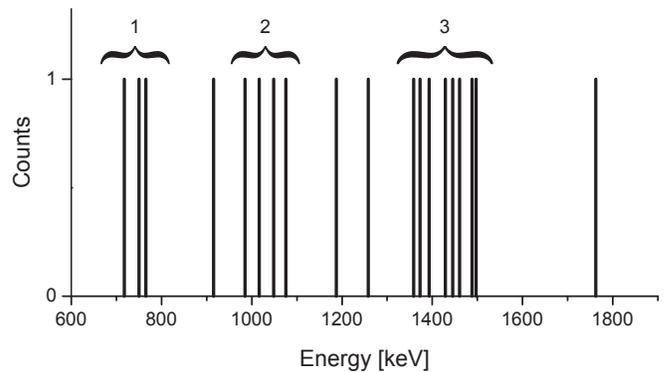}
\caption{\label{fig:Fe} Energy spectrum of $\beta$-delayed protons emitted
from $^{46}$Fe.}
\end{figure}

In most cases the $\beta$-delayed protons had energy large enough to escape from the
chamber. However, 19 such events could be reconstructed. The resulting energy spectrum
is shown in Fig. \ref{fig:Fe}.

Despite the low statistics, there are indications of three lines in the spectrum.
The line located at about $1400$ keV (marked as 3) is compatible with the line at $1457(28)$ keV reported by Dossat et al. \cite{Dossat} which corresponds to the proton kinetic energy of 1425 keV.
The number of counts in this line, corrected by the stopping efficiency of protons,
corresponds to the branching ratio of 3.6(13)\%.
In Ref. \cite{Dossat} this line was assigned the branching of 10(3)\%. Other proton lines observed in Ref. \cite{Dossat} had larger energies, in fact too large to
be reconstructed in the present experiment.
On the other hand, we do see traces of two lines at lower energies (750 keV and 1050 keV) which were not seen by Dossat et al. The corresponding branching ratios are 1.2(7)\% and 1.6(8)\%, respectively. In general the OTPC is more sensitive for low
energy particles than silicon detectors, mainly due to lack of $\beta$ background.

\subsubsection{$\beta 2p$ decay of $^{46}$Fe}

\begin{figure}
\includegraphics[width = 0.95\columnwidth]{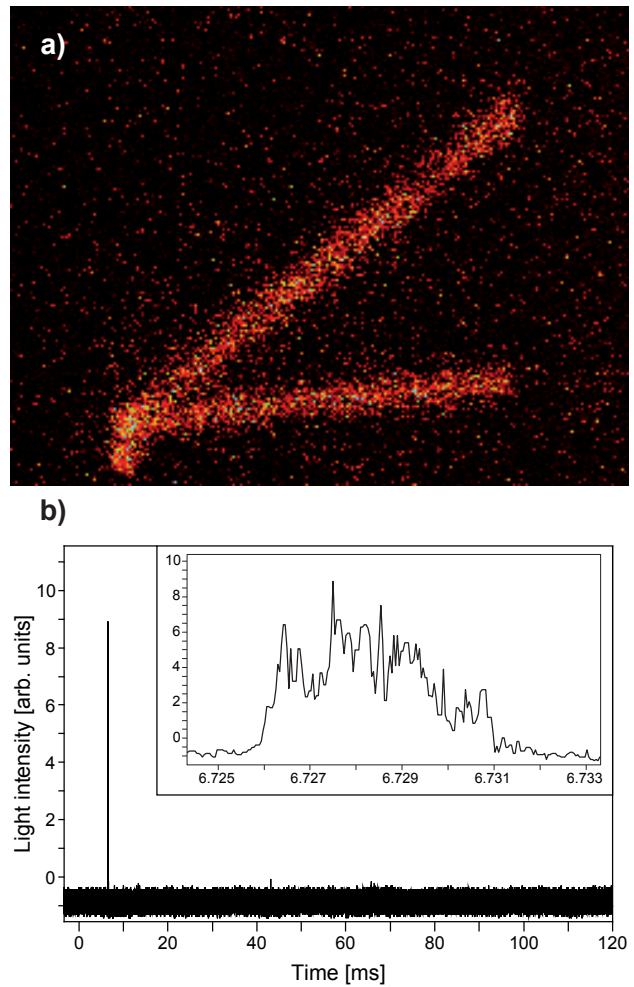}
\caption{\label{fig:2pFe} Decay event showing the $\beta 2p$ emission from $^{46}$Fe.
On the CCD image (a) a short track of the ion entering the chamber from below and two long tracks of particles leaving the detector volume are visible.
The corresponding PMT waveform (b) proves that both particles were emitted at the same
time. Since the ion of $^{46}$Fe stopped very close to the entrance window, the signal
from the implantation, at the zero time, was too small to be visible in this scale.
In the insert the zoomed decay part of the waveform is shown.}
\end{figure}

Among the observed decay events of $^{46}$Fe, there is one clearly showing the simultaneous emission of two high-energy protons. This event, presented in Fig.\ref{fig:2pFe}, provides the first evidence for $\beta$-delayed two-proton emission from this nucleus.
Unfortunately, both protons left the active volume of the OTPC so their energies
could not be reconstructed. From the visible length of both tracks, however, we can
determine the lower limits of their energies. The real length of the two tracks
was evidently larger than 99 mm and 129 mm which for the protons in the OTPC
gas mixture corresponds to an energy larger than 1.67 MeV and 1.96 MeV, respectively.
Thus the energy difference between the proton emitting excited state in $^{46}$Mn
and the final state in the $\beta 2p$ daughter $^{44}$V must have been larger than
3.63 MeV. Taking the $^{44}$V mass excess value $\Delta m = -24.12$~MeV \cite{AME2012}
and the mass excess of the Isobaric Analog State (IAS) of $^{46}$Fe in $^{46}$Mn
as $\Delta m = -7.473$~MeV \cite{Dossat} we obtain the energy difference between these
two states of 2.07~MeV. This means that the two-proton emission proceeded from a
state located more than 1.56~MeV above the IAS state.
One $\beta 2p$ event corresponds to the branching ratio of 0.4(6)\%. This nicely illustrates
the extreme sensitivity of the OTPC detector - one clearly resolved event is sufficient to
claim the observation of a new decay mode.

\subsection{ $^{48}$Ni}

\subsubsection{Production cross-section}

Using the same procedure as in the case of $^{46}$Fe (Eq. \ref{eq:cs}), we can
determine the production cross section for $^{48}$Ni. For the number of identified
ions we use 10, as given by the A1900 ID system, see Fig.\ref{fig:NSCLID}.
The transmission to the final focus of the A1900, $T_1$, was calculated
using the LISE++ code \cite{LISE} according to the procedure described in Ref.\cite{Tarasov2010}, which yielded $T_1=0.34(3)$. The remaining values
were the same as in the case of $^{46}$Fe. The resulting cross section for
the production of $^{48}$Ni in the fragmentation reaction of a $^{58}$Ni
beam at 160 MeV/nucleon on a natural nickel target is $\sigma = (150 \pm 50)$~fb.

In the previous work we had reported the value of $\sigma = (100 \pm 30)$~fb \cite{PomorskiAPP}. The difference arises solely from the $T_1$ coefficient.
In Ref. \cite{PomorskiAPP} it was estimated by using the analytical prediction
of LISE++ and the momentum distribution of Ref. \cite{Tarasov2004}.
Here, we use the more realistic Monte Carlo version of LISE++ and the
momentum distribution given by the Morrissey model \cite{Morrissey},
as recommended by Tarasov et al. \cite{Tarasov2010}. The comparison of our
result with the literature and with the predictions of the EPAX models
is given in Table \ref{tab:Nics}.

\begin{table}
\caption{\label{tab:Nics}Production cross-section for $^{48}$Ni in the
reaction of a $^{58}$Ni beam on a $^{nat}$Ni target. In this work the
beam energy of 160 MeV/nucleon was used, while the value reported
in Ref. \cite{Blank_Ni_disc} was measured at $74.5$ MeV/nucleon.
All values in fb.}
\begin{ruledtabular}
  \begin{tabular}{cccc}
		This work & B. Blank \textit{et al.}\cite{Blank_Ni_disc} & EPAX 2.1 \cite{EPAX2} & EPAX 3 \cite{EPAX3}\\
		\hline
		$150 \pm 50$ & $50 \pm 20$ & $60$ & $20$\\
  \end{tabular}
\end{ruledtabular}
\end{table}

\subsubsection{Half-life and branching ratios}

Nine events of $^{48}$Ni were registered in the OTPC acquisition system.
Two of them did not stop in the active volume of the chamber, so no decay information
could be inferred from them. For six events we did observe the decay accompanied
by emission of protons. In two of them the stopped ion decayed by emission
of a high energy particle, which escaped the active volume of the chamber.
This is interpreted as the $\beta p$ decay of $^{48}$Ni.
Four events represented $2p$ radioactivity of $^{48}$Ni. In two of these
the subsequent decay of $^{46}$Fe ($2p$ daughter of $^{48}$Ni) by $\beta$-delayed
proton emission was also recorded, see Fig. 3 in Ref. \cite{Pomorski2011}.
Finally, in one event the ion was stopped within
the active volume of the chamber but no decay signature was observed during the exposure.
Non-observation of such a signature may indicate that either no protons were emitted
in the decay or that the decay occurred after the decay gate was closed. Both possibilities
are very unlikely, because the $\beta$ daughter, $^{48}$Co is proton unbound \cite{NDS_A48} and
the $^{48}$Ni half-life is much shorter that the decay gate (see below). It could happen, however,
that the decay occurred within the first 100~$\mu$s after the implantation when the OTPC
is still in the low sensitivity mode. In such case, the signal from the emitted proton
would be too weak to be registered. Previously, we have reported two events of this kind \cite{Pomorski2011, PomorskiAPP}. However, after re-analyzing the particle identification
as described in section \ref{chap:ID_fit}, the other event was found to be misidentified.

From the six observed decays of $^{48}$Ni we have determined the half-life using the
maximum likelihood method described in Ref. \cite{LikelihoodT}.
The result is $T_{1/2}= 2.1 ^{+1.4} _{-0.6}$~ms, which is in good agreement with the
value reported in Ref. \cite{DossatNi}.

Based on the observed 4:2 ratio between $2p$ and $\beta p$ decay events, the branching
ratios were determined to be $P_{2p} = 0.7 (2)$ and $P_{\beta p} = 0.3 (2) $ for
the $2p$ and $\beta$-delayed decay channels, respectively.
Combined with the measured half-life this yields the partial half-lives
of $T_{1/2}^{2p} = 3.0 ^{+2.2}_{-1.2}$~ms and $T_{1/2}^{\beta} = 7.0 ^{+6.6}_{-5.1}$~ms
for $2p$ and $\beta$ decay channels, respectively. We note that different branching ratios
were determined for $^{48}$Ni in Ref. \cite{DossatNi}. Out of the four decay events
attributed to $^{48}$Ni, only one was consistent with the $2p$ emission.

\subsubsection{Two-proton radioactivity}

\begin{figure}
\includegraphics[width = \columnwidth]{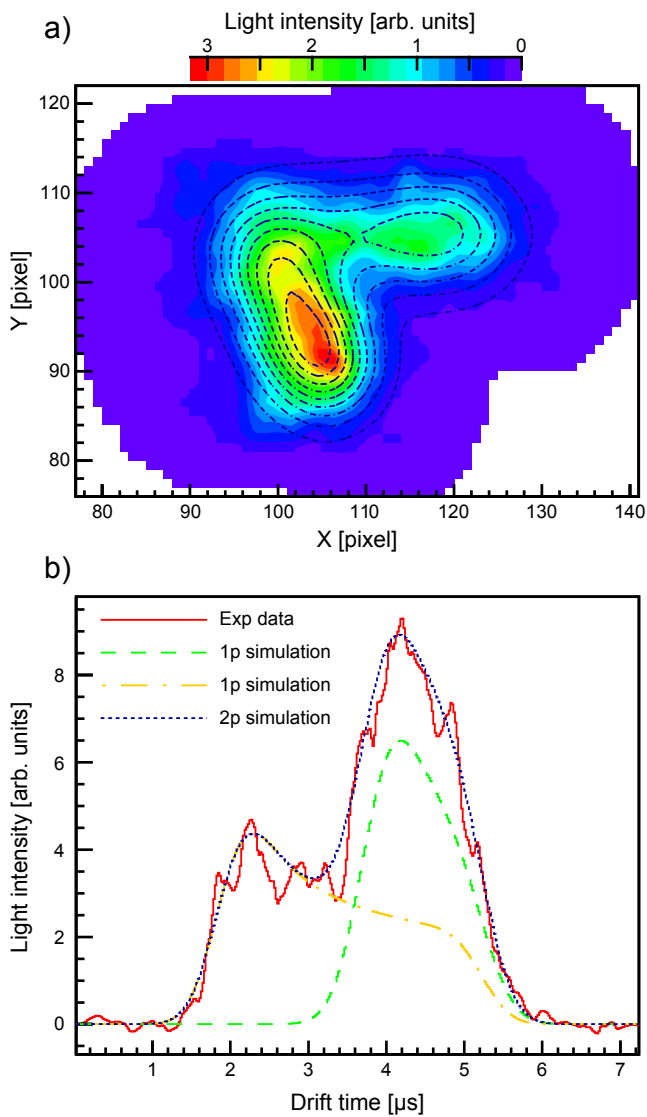}
 \caption{\label{fig:2p} (Color on-line) An example of reconstruction of a $2p$ decay event of $^{48}$Ni. a) On the CCD image represented by a color map the best simulation is shown by contour lines. b) The experimental PMT waveform (red histogram) is shown together with lines representing best fitted traces of individual protons and their sum. Both proton tracks originate at about 5.5 $\mu$s. The track marked by the yellow line represents the proton emitted towards the anode, hence its Bragg peak appears earlier.}
 \end{figure}

Using the procedure described in Sec. \ref{chap:Reconstruction} the four events of $2p$ radioactivity
of $^{48}$Ni were reconstructed. An example of the reconstruction of one event is showed in Fig.\ref{fig:2p}. The results for all four events are presented in Table \ref{tab:2p_res}.
We note that these
results differ somewhat from those published previously in Ref. \cite{PomorskiAPP} where the simplified reconstruction procedure was used.
The weighted average of the $2p$ decay energy is $Q_{2p} = 1.29(4)$~MeV. This value agrees very well with theoretical predictions as shown in Table \ref{tab:2p_E_comp}.

\begin{table}
\caption{\label{tab:2p_res}Results of the reconstruction of the four $2p$ decay events of $^{48}$Ni. $E_1$ and $E_2$ are kinetic energies of both protons, $\theta_{pp}$ is the angle between their momenta, $E_{rec}$ is the daughter recoil energy, and the $Q_{2p}$ is the $2p$ decay energy.}
	\begin{ruledtabular}
		\begin{tabular}{ccccc}
		$E_1$ [keV] & $E_2$ [keV] & $\theta_{pp}$ [deg] & $E_{rec}$ [keV]& $Q_{2p}$ [keV] \\
		\hline
        600(70)  & 645(110) & 66(14) & 37(6)  & 1280 (130) \\
        590(90)  & 635(90)  & 36(7)  & 46(4)  & 1271 (130) \\
        580(60)  & 665(50)  & 51(8)  & 42(4)  & 1287 (80) \\
        645(130) & 680(80)  & 33(17) & 51(7)  & 1373 (160) \\
		\end{tabular}
	\end{ruledtabular}
\end{table}

\begin{table}
\caption{\label{tab:2p_E_comp}Comparison of the determined $Q_{2p}$ value of $^{48}$Ni with theoretical predictions. All values in MeV.}
	\begin{ruledtabular}
		\begin{tabular}{cccc}
			This work &  Brown \cite{Brown} & Ormand \cite{Ormand} & Cole \cite{Cole} \\
		\hline
		$1.29(4)$ &  $1.36 (13)$& $1.29 (33)$ & $1.35(6)$ \\
		\end{tabular}	
	\end{ruledtabular}
\end{table}

\begin{figure}
\includegraphics[width = \columnwidth]{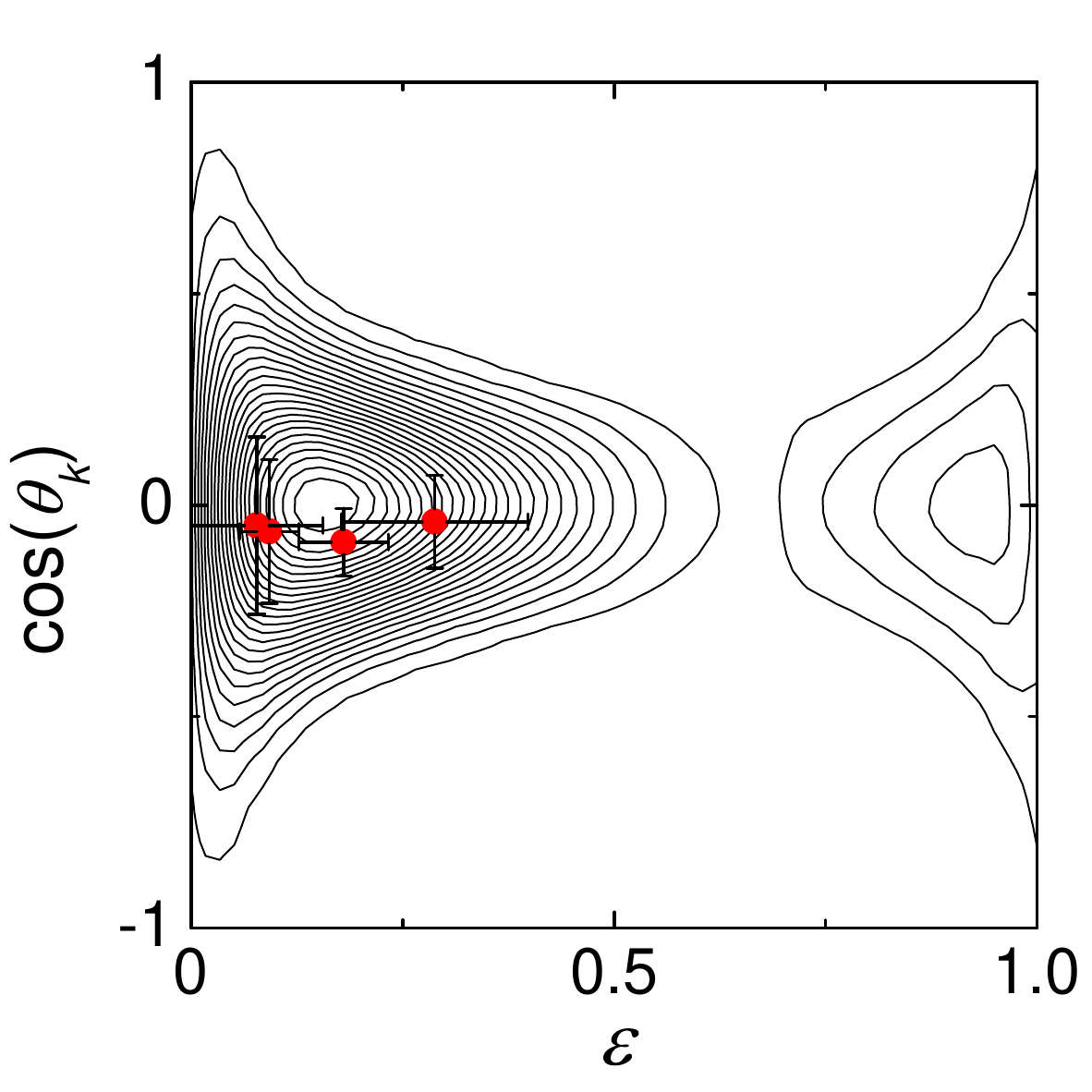}
\caption{\label{fig:48Ni_JacobiT} (Color online) Two-proton momentum correlations from the decay of $^{48}$Ni in the T Jacobi coordinate system, as defined by Eqs. 5,6. It was assumed that the first proton is the one with the lower energy. The opposite assumption would yield the pattern symmetric with respect to the $\cos(\theta_k)=0$ line. The experimental points are superimposed on the contour map of the distribution calculated by the three-body model for the $2p$ decay of $^{45}$Fe \cite{Grigorenko:2010} (adapted with permission from Ref. \cite{PfutznerRMP}.) }
\end{figure}

The measured kinematical data of $2p$ decays can be transformed to the Jacobi coordinate system which is conventionally used in theoretical
description of three-body processes \cite{PfutznerRMP,PfutznerNP152}.
We consider an initial nucleus at rest which decays by emitting
two protons with momenta $\mathbf{k}_1$ and $\mathbf{k}_2$ with the
decay energy of $Q_{2p}$. In the so called T Jacobi coordinate system
we define two Jacobi momenta of two-body subsystems:
\begin{equation}
    {\bf k}_x = \frac{{\bf k}_1 - {\bf k}_2 }{2} , \; {\bf k}_y = {\bf k}_1 + {\bf k}_2 \, .
\end{equation}
Then the complete correlation picture is determined by two parameters,
the energy fraction $\varepsilon$ and the angle $\theta_k$ between the
Jacobi momenta $\mathbf{k}_x$ and $\mathbf{k}_y$:
\begin{equation}
  \varepsilon = \frac{E_x}{Q_{2p}} = \frac{(k_x^2/m_p)}{Q_{2p}} ,\quad \cos(\theta_k)= \frac{(\mathbf{k}_{x} \cdot \mathbf{k}_{y})}{(k_x\,k_y)},
\end{equation}
where $m_p$ is the proton mass and $E_x$ is the energy of protons
with respect to the center of mass of both protons.

Using the data from Tab. \ref{tab:2p_res} we arrive at the Jacobi coordinates
shown in Fig. \ref{fig:48Ni_JacobiT} together with the correlation picture predicted
for the $2p$ decay of $^{45}$Fe by the three-body model \cite{Grigorenko:2010}.
Although the prediction of the model for the case of $^{48}$Ni is missing, it is expected that the distribution will be qualitatively similar to the case of $^{45}$Fe. It has a characteristic feature of two bumps, both centered at $\cos(\theta_k)=0$ with the smaller one at large values of $\varepsilon$. The configuration of both valence protons is assumed to be a mixture of $f^2$ and $p^2$ contributions.
The relative intensity of this smaller bump reflects the contribution of the $p^2$
component \cite{Grigorenko2003}. Obviously more statistics are
needed to establish the experimental distribution for $^{48}$Ni. Presently we can only observe that the four measured points are consistent with the distribution having a maximum at low value of $\varepsilon$ which corresponds to the low relative energy between protons. This is expected if the initial wave function is dominated by protons in the $f^2$ configuration \cite{PfutznerRMP}.

\section{Conclusions}

Using the OTPC detector we have performed proton spectroscopy on nuclei
in the vicinity of the presumably doubly-magic $^{48}$Ni which is presently the most
neutron-deficient corner of the nuclide chart accessed experimentally ($T_z =-4$).
The ions of interest were produced by in-flight fragmentation of a $^{58}$Ni
beam at 160 MeV/nucleon on a natural nickel target and selected from the unwanted
reaction products by the A1900 fragment separator. The ions of $^{48}$Ni were
detected with an average rate of one ion per day with a production cross section
of 150(50)~fb. Such an efficiency and selectivity of the in-flight technique
makes it the method of choice when short-lived very exotic nuclei have to be addressed.

Out of six recorded decays of $^{48}$Ni four decayed by $2p$ radioactivity.
The partial $2p$ decay half-life was determined to be $T_{1/2}^{2p} = 3.0 ^{+2.2}_{-1.2}$~ms.
The reconstruction of the protons tracks yielded the total $2p$ decay energy
of $Q_{2p}=1.29(4)$~MeV, in good agreement with theoretical predictions.
The momentum correlations between protons in these four events are consistent
with the three-body model of $2p$ radioactivity assuming the dominant $f^2$
configuration of the protons. A meaningful comparison of the proton correlations
with the models of $2p$ emission requires much larger statistics. This very
interesting but ambitious task has to wait for the next generation of
radioactive beam facilities.

In addition to $^{48}$Ni, decays of $^{46}$Fe and $^{44}$Cr by $\beta$-delayed
proton emission were also investigated. Although the experimental conditions were
not optimal for this kind of studies, we showed that the OTPC detector can
be successfully used to measure proton spectra, especially at low energy.
The careful reconstruction of tracks left by delayed protons from $^{44}$Cr
yielded the spectrum which shows the same structure as measured previously
by an array of silicon detectors \cite{Dossat} but is much cleaner, having
no contribution from the $\beta$ background. The lack of such background allowed
to identify a new proton line at 740 keV emitted with probability 0.6\% .

Another advantage of the OTPC detector is the accuracy of the branching ratio
determination. The direct counting of the incoming, identified ions and of the
number of events of a specific decay channel leads to the probability value
which accuracy is essentially limited only by statistics.
The possibility to identify unambiguously the decay channel of one event
leads to the extreme sensitivity of the OTPC. This was nicely demonstrated by the
first observation of the $\beta$-delayed two-proton emission from $^{46}$Fe
based on one event, even though both protons escaped from the active volume
of the detector.

\begin{acknowledgments}
We gratefully acknowledge the support of the whole NSCL staff during
the experiment and, in particular, the efforts of the Operations Group to
provide us with the stable, high-intensity beam. This work was supported
by the U.S. National Science Foundation under grant number PHY-06-06007,
by the U.S. Department of Energy under Contracts DE-AC05-00OR22725
and DE-FG02-96ER40983, by the ORNL LDRD Wigner Fellowship WG11-
035 (KM), by the National Nuclear Security Administration under the
Stewardship Science Academic Alliances program through DOE Cooperative
Agreement No. DE-FG52-08NA28552, and by the Polish National Science Center
under contract no. DEC-2011/01/B/ST2/01943.
\end{acknowledgments}


%

\end{document}